\documentclass[%
reprint,
superscriptaddress,
amsmath,amssymb,
aps,
prl,
floatfix,
longbibliography
]{revtex4-2}

\usepackage{times}
\usepackage{amstext}
\usepackage{graphicx}
\usepackage{color}
\usepackage{soul}
\usepackage{MnSymbol}
\usepackage{subfiles}
\usepackage{bm}
\usepackage{hyperref}
\usepackage{layouts}
\usepackage{enumitem}
\usepackage{natbib}
\usepackage{comment}
\usepackage{xcolor}
\usepackage{adjustbox}
\usepackage{float}
\usepackage{gensymb}
\usepackage[abs]{overpic}
\usepackage[normalem]{ulem}

\usepackage[T1]{fontenc}
\usepackage[utf8]{inputenc} 



\begin{document}

\title{Harmonious Color Pairings: Insights from Human Preference and Natural Hue Statistics
}

\author{Ortensia Forni}
\email{ortensia.forni@polytechnique.edu}
\affiliation{Econophysics Lab, Institut Louis Bachelier, 28 Pl. de la Bourse, Palais Brongniart, 75002 Paris, France}
\affiliation{LadHyX UMR CNRS 7646, École Polytechnique, Institut Polytechnique de Paris, 91128 Palaiseau Cedex, France}

\author{Alexandre Darmon}
\affiliation{Art in Research, 33 rue Censier, 75005 Paris, France}

\author{Michael Benzaquen}
\affiliation{Econophysics Lab, Institut Louis Bachelier, 28 Pl. de la Bourse, Palais Brongniart, 75002 Paris, France}
\affiliation{Capital Fund Management, 23 Rue de l’Université, 75007 Paris, France}

\begin{abstract}
While color harmony has long been studied in art and design, a clear consensus remains elusive, as most models are grounded in qualitative insights or limited datasets. In this work, we present a quantitative, data-driven study of color pairing preferences using controlled hue-based palettes in the HSL color space. Participants evaluated combinations of thirteen distinct hues, enabling us to construct a preference matrix and define a combinability index for each color. Our results reveal that preferences are highly hue dependent, thereby challenging classical color harmony theories. Yet, when averaged over hues, statistically meaningful patterns of aesthetic preference emerge, with certain hue separations perceived as more harmonious. Strikingly, these patterns align with hue distributions found in natural landscapes, pointing to a statistical correspondence between human color preferences and the structure of color in nature. Finally, we analyze our color-pairing score matrix through principal component analysis, which uncovers two complementary hue groups whose interplay underlies the global structure of color-pairing preferences.
Together, these findings offer a quantitative framework for studying color harmony and its potential perceptual and ecological underpinnings.
\end{abstract}

\date{\today}

\maketitle

\section{Introduction} 

Early efforts to understand the nature of color can be traced back to Antiquity with Aristotle’s chromatic theories~\cite{taschen2025}. This said, throughout much of history, color was primarily associated with symbolic and religious meanings. In  Western civilizations, for example, saturated colors and polychromy were often associated with superficiality, vulgarity, or foreignness, while whiteness was imbued with connotations of purity and reassurance. A scientific turn occurred in the seventeenth century  with the emergence of a more intellectual and systematic approach to color, exemplified by the creation of various color charts and tables, such as Richard Waller’s \textit{Catalogue of Simple and Mixt Colours}~\cite{waller1686}. A major scientific milestone was reached in 1704 with Newton’s famous treatise, \textit{Opticks}~\cite{newton1704}.

Since then, the ambition to build consistent and rigorous color theories has continued to grow, drawing interest from a wide range of disciplines including mathematics, psychology, neuroscience, philosophy, and design, to name a few. A considerable body of research has focused on the construction of well-defined color spaces, whether through strictly mathematical frameworks~\cite{weinberg1976geometry,kuehni2001color,bujack2022non,berthier2022quantum, BERTHIER2021102562} or more artistically driven approaches~\cite{munsell1915atlas}. One particularly prominent area of color theory that has attracted sustained interest over the centuries is the study of color harmony~\cite{bezold1876,itten1961,granville1987color,matsuda1995}. Goethe~\cite{goethe1840}, Chevreul~\cite{chevreul1855}, Ostwald~\cite{ostwald1932}, among others, first sought to capture the essence of harmonious color pairings by linking aesthetic appreciation to the perceptual distance between colors. More recently, data-driven approaches have gained traction. Some studies have focused on color emotion~\cite{ou2004study_part1, ou2004study_part3, valdez1994effects, ou2018universal, ou2004study_part2}, aiming to identify correlations between specific hues and the emotions they trigger. Other works have investigated color preferences in terms of biological adaptations, highlighting marked gender differences~\cite{ling2007new, hurlbert2007biological}, or through the \textit{Ecological Valence theory} ~\cite{palmer2010ecological}, which links color preferences to associations with objects perceived positively or negatively. Additional studies have examined how preferences vary with age{~\cite{dittmar2001changing, taylor2013color}}, art education{~\cite{oliveira2025color}}, or seasonality{~\cite{schloss2017seasonal}}. Building on these perspectives, researchers have also sought to develop generalizable models of color harmony using machine learning techniques~\cite{lu2014discovering,xu2025} and to build computational toolkits for extracting image features relevant to aesthetic judgments{~\cite{redies2025toolbox}}. Complementing these scientific developments, a distinctly historical-cultural line of work has emphasized the enduring importance of colours across history and the ways in which the social, moral, and symbolic values attributed to them evolve over time, shaped by historical contexts and collective practices {\cite{Histoire_des_couleur_Pastoureau, pastoureau2005petit}}.

In spite of these developments, most approaches to the color pairing problem still lack a truly quantitative foundation. Landmark perception-based studies such as those by Moon and Spencer~\cite{moon1944aesthetic} or Granger~\cite{granger1955experimental} have produced widely cited harmony principles, but were based on a very limited number of survey participants. A more recent and ambitious study was conducted by Nemcsics~\cite{nemcsics2009experimental} in the Coloroid space among Technology and Economics students at Budapest University. 
However, the high level of complexity in the study—both in the geometrical patterns presented to participants and in the selection of colors combining hue, saturation, and lightness—may limit the interpretability of the results.

In summary, to this day, there remains no clear consensus on the principles underlying color harmony~\cite{o2010colour}, thus leaving the topic open to debate and inquiry. 
Our approach here is inspired by the work of Lakhal \textit{et al.}~\cite{Lakhal_2020}, on the structural complexity of black-and-white images, in which some of us provided evidence for a certain level of universal quantitative criteria for aesthetic judgment, see also~\cite{granger1952objectivity}. Interestingly, the preferred level of complexity correlated strongly with that found in natural images, suggesting that what we find aesthetically pleasing may be shaped by what we are most frequently exposed to. Related work in the color domain has similarly connected aesthetic preferences to statistical regularities in natural images and paintings, pointing to a degree of universality in preferred chromatic compositions{~\cite{montagner2016statistics, nakauchi2022universality, nascimento2021naturalness, mcdermott2012uniform}}. In line with this approach, we conduct a large-scale survey targeting a sufficiently diverse participant panel and grounded in simple evaluation procedures, aiming both to gain insights into preferences for color pairings and to compare these preferences with color combinations found in natural images.

The remainder of this paper is structured as follows. We first introduce our survey methodology, which is based on direct comparisons of simple color pairs. Next, we analyze the resulting data and propose a metric that quantifies how harmoniously each color pairs with others. We then explore how this metric aligns both with absolute color preferences reported in independent studies and with hue distributions observed in natural images. By averaging across hues, we identify angular distances between colors that are consistently preferred or rejected, and compare these perceptual trends to signals found in natural landscapes. Finally, we examine how different color groups contribute to the emergence of configurations of appreciation and rejection.\\

\begin{figure}
    \centering
    \includegraphics[width=0.5\linewidth]{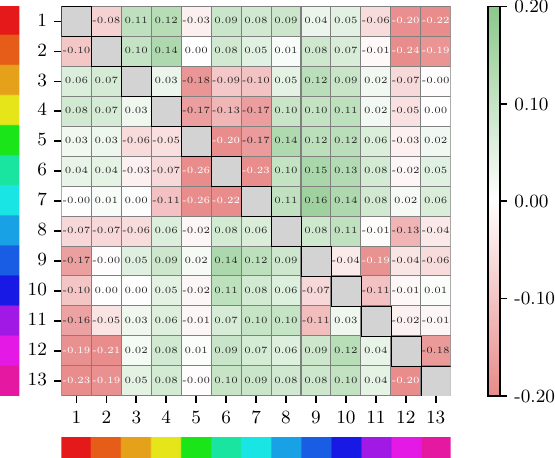}
    \caption{Preference matrix S computed from the survey results and Eq.~\eqref{eq:score}.}
    \label{fig:matrix}
\end{figure}

\section{Survey}
We conduct a large-scale survey in which participants are asked to select their three most and three least preferred color combinations from sets of predefined color palettes, described in details in the following sections. Our study operates within the HSL (Hue, Saturation, Lightness) color space~\cite{joblove1978color}, focusing exclusively on the hue component while keeping saturation and lightness fixed at $\text L = 0.5$ and $\text S = 0.8$~\footnote{Note that the HSL color space was chosen because it allows hues to be easily isolated, which would not be possible in other commonly used spaces such as RGB, which requires the use of all three channels. Interestingly enough, the choice of color space does not appear to affect the results on color preferences, see e.g.~\cite{ling2007new}. Most importantly, hue preferences have been found to remain largely unaffected by variations in lightness and saturation~\cite{granger1955experimental}.}. This choice also supports our large-scale, quantitative objective: by focusing on hue—which is less affected by between-device variability than saturation and lightness—we were able to conduct the survey online and recruit a large, heterogeneous participant pool. Hues are represented as angular values on the HSL color wheel~\cite{parkhurst1982invented}, with red as reference color positioned at 0°, following the usual convention.
We sample 18 equally spaced hues at $20^{\circ}$ intervals, starting from 0°, providing a reasonably fine coverage of the full hue wheel while keeping the set manageable for a preference task; then we remove five of them (corresponding to 80°, 100°, 140°, 260°, and 340°) to keep only clearly distinguishable colors (verified across different display devices). Indeed, the HSL color space lacks perceptual uniformity~\cite{lissner2011toward}, as some colors comprised within large angular ranges (such as greens) cannot be unmistakably separated by the eye.  
In the following we refer to the 13 retained colors with indices \( i \in \{1, \dots, 13\} \), with 0° red corresponding to $i=1$. 
We then generate 13 sets of 12 color pairs by taking each color $i$ in turn as the reference color, and pairing it with each of the remaining 12 colors $j$, hereafter referred to as the counterpart colors. The color pairs are presented as checkerboards consisting of 8$\times$8 square grids, with the reference and counterpart colors distributed in equal amounts, though randomly in space to avoid recognizable patterns that might bias participants’ responses. An example of set with $\text H=200$° blue as reference color is shown in the Supplemental information (Fig.~\ref{fig:colorpalettes}). The survey is structured as follows: for each of the 13 sets, a question is created in which participants are shown, in a random order, one of the sets. Participants are then asked to select the three most and three least harmonious color combinations. To limit survey duration and maintain engagement, each participant was asked to complete 6 randomly assigned questions
(out of 13), rather than the full set. The subset of questions varied across participants, ensuring a comparable number of responses for each of the 13 reference hues.
The survey was conducted \textit{via} the Qualtrics platform~\cite{qualtrics}. We collected 346 responses from colleagues at École Polytechnique and CFM (France), OIST (Japan), as well as volunteers across Europe  who participated without financial incentives. The sample included both male and female participants aged between 20 and 65, mostly with a high level of education but from diverse academic backgrounds.\\

\section{Results}

For each reference color $i$ and counterpart color $j$, we define $f^\text B_{ij}$ as the frequency with which the color combination $(i, j)$ was selected among participants as one of the three most harmonious combinations, normalized to the number of survey responses per question. Analogously, $f^\text W_{ij}$ is the normalized frequency with which $(i, j)$ was selected as one of the three least harmonious combinations.  
A high value of $f^\text B_{ij}$ suggests that color $i$ pairs well with color $j$, though a low value does not necessarily imply a negative judgment; it may simply reflect infrequent selection among the top three. Conversely, a high $f^\text W_{ij}$ suggests unpleasantness while a low value does not imply positive judgment.
We thus define the overall score of each color pair as: 
\begin{equation}\label{eq:score}
    \text S_{ij} = f^\text B_{ij} - f^\text W_{ij} .
\end{equation}
The resulting score matrix $\text S$ is shown in Fig.~\ref{fig:matrix}, where the $i^{\text{th}}$ row corresponds to responses to the question in which color $i$ was used as the reference.  Note that the rows of the matrix naturally sum to zero (participant answers for each question were only retained if they provided all three best and three worst choices). 
Remarkably, $\text S$ reveals an underlying structure, with discernible clusters of color combinations that are perceived as either harmonious or inharmonious. The nature of these clusters is examined in greater depth later in the paper.

\begin{figure}
    \centering
    \includegraphics[width=0.5\linewidth]{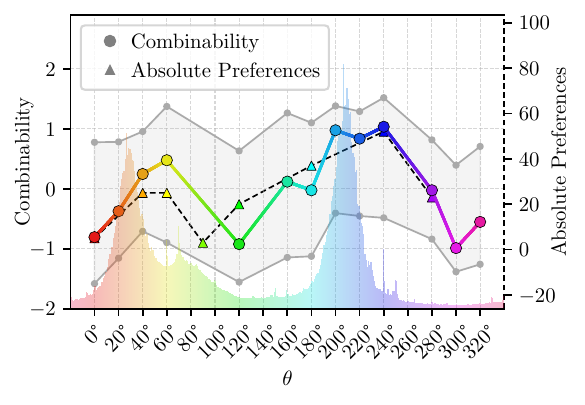}
    \caption{Combinability index from Eq.~\eqref{eq:combin} (circular markers, solid line), absolute preferences from~\cite{palmer2010human} (triangular markers, dashed line), and hue distribution of 12,000 natural landscape images (background histogram), as a function of hue angular values $\theta$. The  solid gray lines reflect the contribution of each color to both the best-only and worst-only indices (see main text).}
    \label{fig:Overlap}
\end{figure} 

Reassuringly, the score matrix appears to be quasi-symmetric (see Fig.~\ref{fig:eigenvectors}a), indicating that the appreciation or dislike of a color combination does not strongly depend on the specific set in which it is presented to survey participants. In other words, interchanging the reference and counterpart colors leaves the results largely unchanged to first order. 
That said, the matrix is not perfectly symmetric, and the residual asymmetry carries meaningful information. While the row-wise normalization only allows to capture relative preferences for specific color pairs, the columns reflect judgments of absolute color combinability—that is, the ability of a given color $j$ to harmonize with all other colors across the set.
For example columns 8, 9, or 10--corresponding to 200°, 220°, and 240° (shades of blue)--are composed of a majority of positive scores. This indicates that these hues tend to be selected more frequently as part of the best than worst combinations. On the contrary columns of red and purple (columns 1, 12 or 13) carry mostly negative values, suggesting that these colors tend to produce less harmonious combinations when paired with others.\\

\section{Combinability index}

To quantify how harmoniously a color tends to combine with others, we define, for each color $j$, the Combinability index  as the sum of its pairing frequencies with all the other colors, across all questions:
\begin{equation}\label{eq:combin}
    \text{C}(j) = \sum_{ i\neq j} \text S_{ij} .
\end{equation}
The combinability indices for the 13  colors of our study are plotted in Fig.~\ref{fig:Overlap}. 
Following the shades of blue (200°, 220°, and 240°) discussed above, yellow (60°), closely followed by orange (40°), exhibit the highest Combinability indices.
Conversely, red (0°), green (120°) and purple (300°) tend to produce combinations that are more frequently perceived as inharmonious. 
The top and bottom gray lines delimiting the shaded area represent, respectively, the sum $\text{C}^\text B(j)$ and $\text{C}^\text W(j)$ of the best-only frequencies $f^\text B_{ij}$
and the worst-only frequencies $f^\text W_{ij}$ for each color $j$. Interestingly, although certain colors like 180° cyan ($j=7$) have a Combinability index close to zero, they show substantial values of $\text{C}^\text B$ and $\text{C}^\text W$, indicating that such hues are frequently selected by participants—appearing as often in the most appreciated combinations as in the least.

Building on the intuition that combinability might be linked to absolute color preference, we compare our results with findings from the Berkeley Color Project~\cite{palmer2010human}, which investigated human preferences for eight highly saturated individual colors (triangular markers and dashed line in Fig.~\ref{fig:Overlap}). 
Remarkably, individual hue preferences closely mirror the Combinability index, suggesting that a color’s ability to form harmonious combinations aligns with its overall aesthetic appeal.

Finally, leveraging previous studies that have highlighted a strong correlation between human aesthetic preferences and structures found in nature~\cite{Lakhal_2020, PhysRevLett.110.018701, tolhurst1992amplitude}, we analyzed a dataset of 12,000 landscape images~\cite{kaggle_12000} spanning various natural biomes, including coasts, deserts, forests, glaciers, and mountains. The count of hue occurrences in the entire database is shown as a background histogram in Fig.~\ref{fig:Overlap}. Strikingly, the peaks and valleys of the distribution match remarkably well that of the Combinability index and absolute preferences (maximum for blue and orange-yellow tints, minimum for greens and purples), 
suggesting that our aesthetic preferences may be influenced by the colors we have most frequently been exposed to.
To test the robustness of our findings and avoid potential biases in image selection, we also measured the hue distribution using two additional non-overlapping datasets of 4,319 and 15,501 natural images respectively~\cite{kaggle_4000, zhou2017places}\footnote{These latter correspond to the validation split of the Places database~\cite{zhou2017places}, restricted to 31 categories associated with natural scenes.}; the results were found to be virtually identical.\\

\section{Preferred combinations}

\begin{figure}
    \centering
    \includegraphics[width=0.5\linewidth]{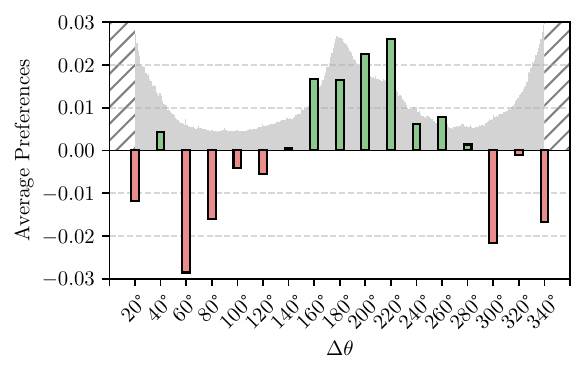}
    \caption{Average preferences for color pairs as function of their angular distance $\Delta\theta$ on the hue wheel (see also Fig.~\ref{fig:wheel_errorbar}). The background gray histogram represents the distribution of angular distances between hues found in natural images. The maximum angular distance between hues is $\Delta\theta=180^\circ$.}
    \label{fig:nature}
\end{figure}

Classical color harmony theories, such as those of Moon and Spencer~\cite{moon1944aesthetic} or Itten~\cite{itten1961}, tend to rely on a strong assumption of hue independence, meaning that preferences for a given hue pair depend only on the angular distance separating the two colors on the hue wheel, regardless of their absolute positions. To examine the relative positions of the combined colors $(i,j)$, we rotate the hue wheel in each set such that the reference color is set at $0^\circ$ (see Fig.~\ref{fig:all_Supplemental information}), and assign to each comparison color a value function of its angular distance from the reference.
Our findings challenge this hue-independence assumption to a large extent, as discussed further below. Nevertheless, in order to confront our findings with existing theories, we compute, for each angular distance, the average selection frequency across all reference colors. 
The results are shown in Fig.~\ref{fig:nature}, where green bars indicate preference and red bars signify dislike, with $180^\circ$ corresponding to the maximum angular distance between two hues. Higher preferences are observed in the \textit{contrast} region (between $160^\circ$ and $220^\circ$), flanked by regions of lower preference at smaller angular distances. Such global preference for contrast aligns rather well with the universal harmony model proposed by Moon and Spencer\cite{moon1944aesthetic}, which remains influential today \footnote{Although Moon and Spencer used the Munsell color system~\cite{munsell1915atlas}, while we rely on HSL, the hue wheels are broadly comparable, and our conclusions do hold to first order.}. However, the other preference regions they proposed—coined \textit{similarity} and located near the reference color—are not consistently supported by our hue-averaged data.
The universality of such theories is further challenged when examining the standard deviation at each angular distance, shown as error bars in Fig.~\ref{fig:wheel_errorbar} (see Supplemental information). The substantial variability observed reflects pronounced differences across individual hue preference wheels (see Fig.~\ref{fig:all_Supplemental information}). Consequently, the assumption of hue independence cannot be reasonably upheld. For instance, the preference wheels for green (e), cyan (g), and light purple (l) in Fig.~\ref{fig:all_Supplemental information} display markedly distinct patterns~\footnote{Also note that, contrary to Moon and Spencer’s theory~\cite{moon1944aesthetic}, the preference wheels are not symmetric with respect to angular distance--preferences at $\Delta\theta$ are different from those at $-\Delta\theta$. See, for instance, cyan (g) in Fig.~\ref{fig:all_Supplemental information} for a striking illustration.}.

To compare our results with hue combination occurrences in natural images, we identify the dominant hue of each landscape in our dataset using a Gaussian kernel density estimation~\cite{wkeglarczyk2018kernel}. For each image, we then count the number of pixels falling at each angular distance from the dominant hue. Summing these counts over the entire dataset yields the gray histogram shown in Fig.~\ref{fig:nature}. Note that angular regions between 0°–20° and 340°–360° were excluded from the analysis, as they are artificially overrepresented: since the dominant hue is always aligned to 0°, nearby shades of the same tint are, by construction, more likely to appear. 
At sufficiently large angular distances from the reference color, dominant hues in natural scenes are most often separated by approximately 180°, suggesting that natural environments are largely characterized by strong color contrasts.
The overlay of the empirical preferences and the natural color distance histogram reveals a compelling alignment: the most appreciated color pairs in our survey tend to coincide with those that occur more frequently in nature, while the least appreciated combinations correspond to rarer configurations in natural landscapes. This suggests that our aesthetic preferences for color pairings may be shaped—at least in part—by repeated exposure to common visual patterns in our environment. This finding strongly  echoes  earlier results by some of us~\cite{Lakhal_2020} on black-and-white image complexity, where aesthetic appeal was likewise found to correlate with structural patterns prevalent in nature. Note that the alignment between preferences and natural color pair occurrences can also be analysed on a hue-by-hue basis (see Fig.~\ref{fig:all_Supplemental information}).

\begin{figure*}[t]
    \centering
    \includegraphics[width=\linewidth]{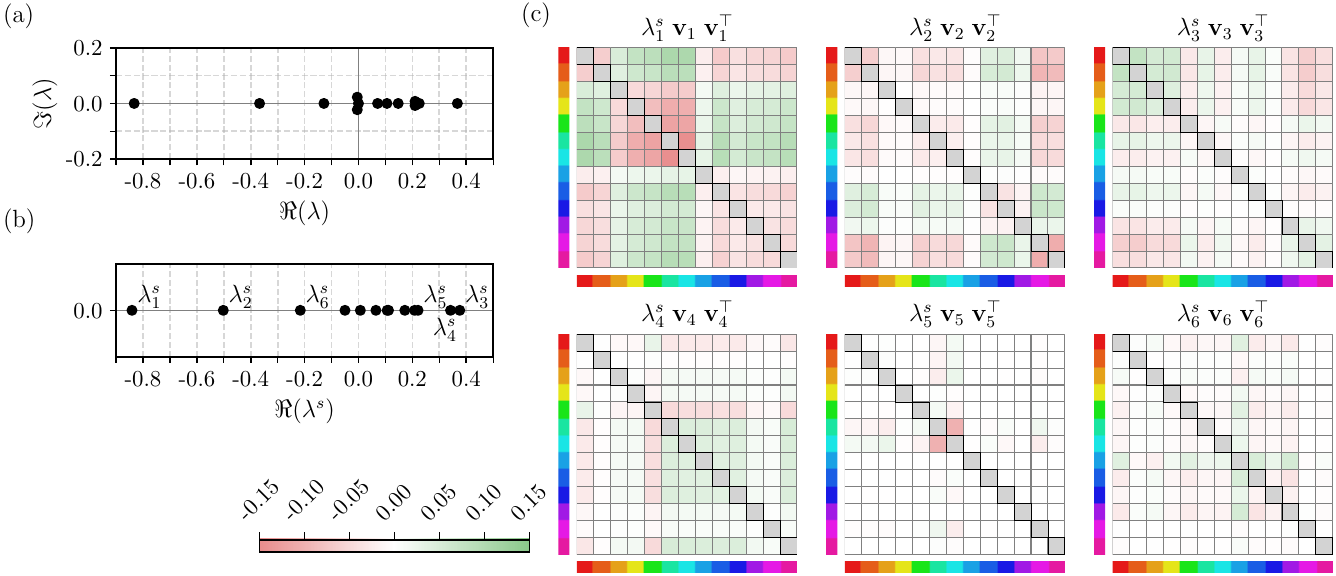}
    \caption{(a) Eigenvalues of the score matrix S  on the complex plane. (b) Eigenvalues of the symmetrized score matrix S$^s= \frac{1}{2}(\text S + \text S^\top)$. (c) Outer products $\lambda_i^\text s \bm{\text v_i} \bm{\text v_i}^\top$ for the six eigenvalues $\lambda_i^{\text s}$ with the highest absolute values, where $\bm{\text v_i}$ denotes the eigenvector associated with the $i^{\text{th}}$ eigenvalue.}
    \label{fig:eigenvectors}
\end{figure*}

\begin{figure}
    \centering
    \includegraphics[width=0.45\linewidth]{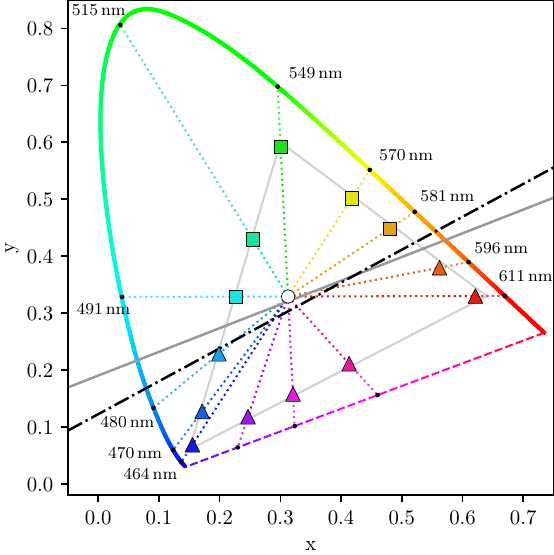}
    \caption{Representation of our 13 sampled hues in the CIE 1931 xy chromaticity diagram. The outer curved boundary represents the spectral locus and the light-gray triangle corresponds to the sRGB gamut{~\cite{anderson1996proposal, susstrunk1999standard}}. Hues belonging to group 1 and group 2 are plotted as squares and triangles respectively, and dotted lines project each color to its dominant wavelength on the spectral locus. The two separating lines are obtained using hard-margin Support Vector Machine (SVM){~\cite{cortes1995support}}. The gray one is the standard hard-margin solution, which maximizes the distance to the nearest points of both groups, while the black dash-dotted line is obtained by assigning a weight to the two support vectors on each side that, intuitively, encodes how strongly a color belongs to its group and is defined as
$\omega_i = \big( \sum_{j:\, \mathrm S^{s}_{ij}<0} |\mathrm S^{s}_{ij}| \big)\big/ \big( \sum_{p,q:\, \mathrm S^{s}_{pq}<0} |\mathrm S^{s}_{pq}| \big)$ .}
    \label{fig:gamut}
\end{figure}

To better understand the structure of color pairing preferences—and to assess whether certain colors or groups of colors contribute more significantly to the overall signal—we analyze the symmetric part of the score matrix, defined as $\text S^s = \frac{1}{2}(\text S + \text S^\top)$, and study its principal components.
As previously noted, the original score matrix $\text S$ exhibits a near-symmetric structure, which is quantitatively supported by the proximity of its eigenvalues to the real axis, see Fig.~\ref{fig:eigenvectors}a. This observation justifies our focus on the symmetric component $S^s$, the eigenvalues of which are plotted in Fig.~\ref{fig:eigenvectors}b. In Fig.~\ref{fig:eigenvectors}c, we display the outer product matrices $\lambda_i^{\text s} \bm{\text v_i} \bm{\text v_i}^\top$ corresponding to the six eigenvalues $\lambda_i^{\text s}$ with the highest absolute values  (see Fig.~\ref{fig:eigenvectors}b), where $\bm{\text v_i}$ denotes the eigenvector associated with the $i^{\text{th}}$ eigenvalue.
The first matrix, associated with the largest eigenvalue, reveals a clear emergence of positive and negative clusters, indicating a division of the hue wheel into two main groups: one spanning from orange to cyan (group 1), and the other from blue to dark orange (group 2). Colors from group 1 tend to combine harmoniously with those from group~2, but not with others within their own group, and vice versa. This structured clustering highlights which segments of the hue spectrum predominantly drive the universal appreciation seen in the contrast region of Fig.~\ref{fig:nature}, as well as the adjacent zones of disfavor.\\
{Interestingly enough, when our color groups are represented in the standard CIE 1931 xy chromaticity diagram~\cite{schanda2007colorimetry} (see Fig.~\ref{fig:gamut}), they are found to be linearly separable. One can also clearly see that while group 1 includes colors from the central region of the visible spectrum, group 2 is composed of colors from two extremes of such spectrum, together with the purples (which correspond to a mix of blue and red). Finally note that the decision boundaries shown in Fig.~\ref{fig:gamut} 
pass close to white, at the center of the gamut.}\\
Further structure is visible in the subsequent components. In particular, the second matrix exhibits a more localized negative signal, concentrated among hues from light purple to dark orange (bottom right red cluster), indicating internal incompatibilities within this range. Additionally, the fifth matrix reveals another pronounced negative contribution, pointing to poor combinability between green and cyan.\\

\section{Concluding remarks} 

Let us summarize the main contributions of this work. We designed and conducted a large-scale survey in which participants were asked to select the most and least appealing color combinations from carefully curated sets. Our first finding was that certain hues—such as blue and yellow—consistently form more harmonious pairings with other colors. We then introduced a combinability index to quantify this behavior, and showed that it correlates well with absolute color preferences reported in an independent study~\cite{palmer2010human}. More notably, this index also aligns closely with the distribution of hue occurrences in natural landscapes.
Shifting focus from individual hues to angular distances between paired colors, we uncovered a robust preference for combinations in the contrast region, centered around complementary hues. This pattern mirrors color distributions commonly found in nature, suggesting that aesthetic appeal may be shaped, at least in part, by frequent exposure to naturally occurring visual stimuli (see also~\cite{Lakhal_2020}).
That said, our results do not fully align with the universal view of classical color harmony theories~\cite{itten1961, granville1987color, matsuda1995, goethe1840, chevreul1855, ostwald1932}, which assumes that harmony depends mainly on fixed color distances (e.g. simple angular relations) rather than on the absolute colors involved. While contrast is favored on average, preferences vary strongly with the specific hues paired, so fixed-distance rules do not consistently predict human preferences.
Finally, a principal component analysis of the score matrix provided further insight, revealing clusters of hues that tend to combine either particularly well or poorly with others, thus offering a more nuanced understanding of the structure underlying human color pairing preferences.\\

\section{Limitations of the study} 

To simplify the interpretation of color combinability and offer an accessible entry point into the question of color harmony, we deliberately limited our analysis to variations in hue only. This choice is also well suited to a large-scale online survey, as hue is comparatively more robust than saturation and lightness to differences across display devices. While this approach offers clarity and tractability, it overlooks the potentially rich interactions with saturation and lightness. A natural direction for future work would be to extend the study to full-color combinations across all HSL dimensions, and to examine how such variations relate to patterns found in natural imagery.
Another important avenue lies in broadening the participant panel—not only in size but also in cultural and demographic diversity—to test the generalizability of our findings across different populations.
Finally, our use of the HSL color space, though convenient for isolating hue, deserves further discussion. HSL suffers from perceptual non-uniformity and device dependence, limiting its fidelity in capturing human color perception. Future research could explore alternative, perceptually uniform color spaces such as CIELAB (CIE $\mathrm{L}^{*}\mathrm{a}^{*}\mathrm{b}^{*}$ defined by the \textit{Commission Internationale de l'Éclairage}) and OKLAB~\cite{ottosson2020perceptual}. However, this comes with its own challenges: in such models, hue is no longer a single scalar dimension but arises from interactions between multiple components, complicating direct comparisons.\\

\section{Acknowledgments}
We are deeply grateful to all the participants of our survey, whose contribution was foundational to this work. We also thank Jean-Philippe Bouchaud, Pierre Bousseyroux, Samy Lakhal, Elia Moretti and Mirko Polato for fruitful discussions. This research was conducted within the Econophysics \& Complex Systems Research Chair, under the aegis of the Fondation du Risque, the Fondation de l’Ecole polytechnique, the Ecole polytechnique and Capital Fund Management.

\clearpage
\bibliography{refs}

\clearpage

\section*{Supplemental information}

\renewcommand{\thefigure}{S\arabic{figure}}
\setcounter{figure}{0}

\makeatletter
\renewcommand{\theHfigure}{S\arabic{figure}}
\makeatother

\begin{figure}[h!]
    \centering
    \includegraphics[width=0.5\linewidth]{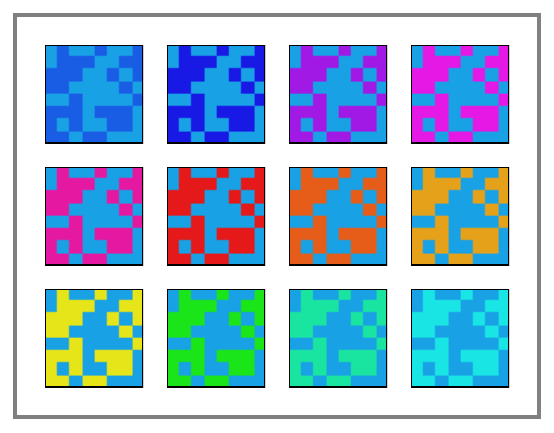}
    \caption{Example of a survey set with  $\text H =200$° as reference color.}
    \label{fig:colorpalettes}
\end{figure} 
\vspace{-0.3cm}
\begin{figure}[h!]
    \centering
    \includegraphics[width=0.5\linewidth]{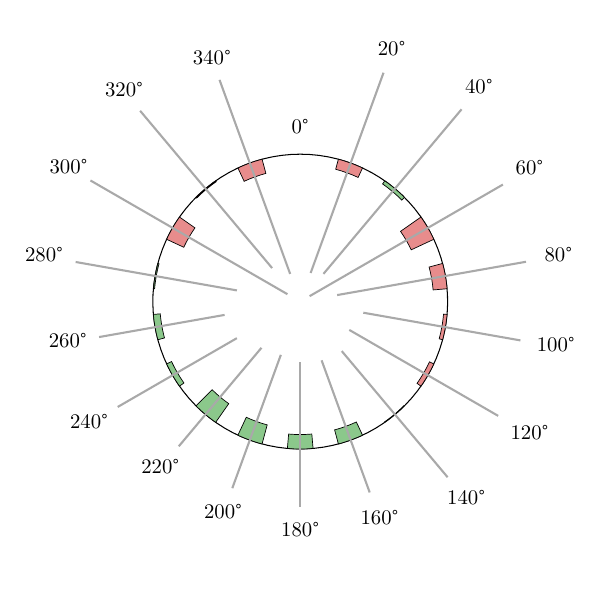}
    \caption{Average preferences for color pairs as function of their angular distance--as in Fig.~\ref{fig:nature}--presented on a wheel with  error bars signifying standard deviations. Green corresponds to positive values, while red indicates negative ones.}
    \label{fig:wheel_errorbar}
\end{figure} 


\begin{figure*}[h!]
    \centering
    \includegraphics[width=\linewidth]{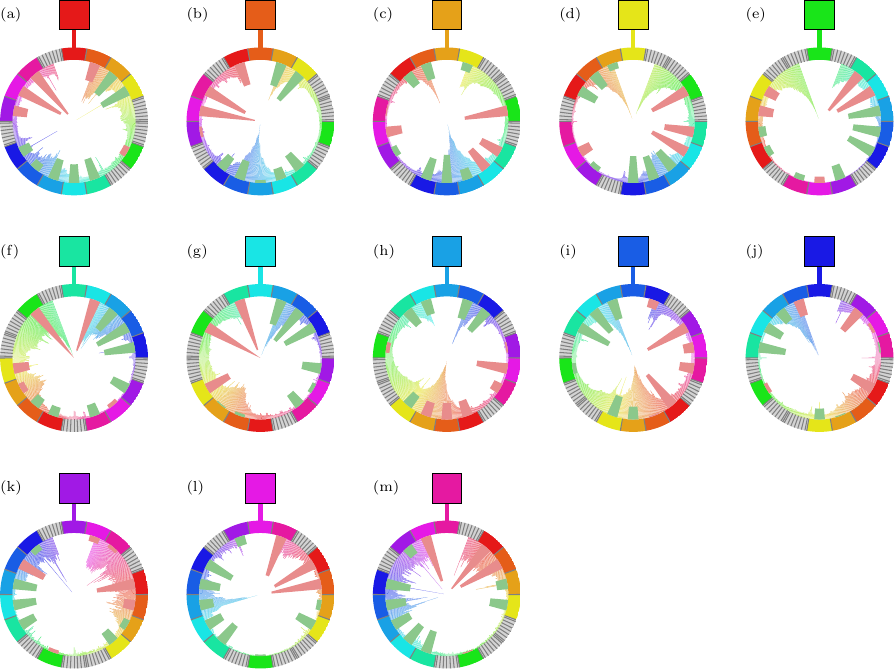}
    \caption{Color pair preferences for each of the 13 hues considered. Green corresponds to positive values, while red indicates negative ones. The background histograms show the distribution of angular distances relative to the reference hue, calculated over subsets of natural images where this hue appears as the dominant color.}
    \label{fig:all_Supplemental information}
\end{figure*}


\end{document}